\documentclass[manuscript]{aastex63}

\usepackage{amsmath} 
\usepackage{amsbsy}
\UseRawInputEncoding 
\accepted{2023 March 14}
\submitjournal{ApJ}

\shorttitle{Sample article}
\shortauthors{Wu et al.}

\begin{document}

\title{Scaling anisotropy with stationary background field in the near-Sun solar wind turbulence}
\correspondingauthor{Honghong Wu}
\email{honghongwu@whu.edu.cn}

\author{Honghong Wu}
 \affiliation{School of Electronic Information, Wuhan University, Wuhan, People's Republic of China}

\author{Jiansen He}
\affiliation{School of Earth and Space Sciences, Peking University, Beijing, People's Republic of China} 
 
\author{Shiyong Huang}
\affiliation{School of Electronic Information, Wuhan University, Wuhan, People's Republic of China} 

 \author{Liping Yang}
\affiliation{SIGMA Weather Group, State Key Laboratory for Space Weather, National Space Science Center, Chinese Academy of Sciences,Beijing, People's Republic of China} 

\author{Xin Wang}
\affiliation{School of Space and Environment, Beihang University, Beijing, People's Republic of China} 

\author{Zhigang Yuan}
\affiliation{School of Electronic Information, Wuhan University, Wuhan, People's Republic of China} 


 


\begin{abstract}    
The scaling of magnetic fluctuations provides crucial information for the understanding of solar wind turbulence. However, the observed magnetic fluctuations contain not only turbulence but also magnetic structures, leading to the violation of the time stationarity. This violation would conceal the true scaling and influence the determination of the sampling angle with respect to the local background magnetic field. Here, to investigate the scaling anisotropy, we utilize an easy but effective criterion $\phi<10^\circ$ to ensure the time stationarity of the magnetic field, where $\phi$ is the angle between the two averaged magnetic fields after cutting the interval into two halves. We study the scaling anisotropy using higher-order statistics of structure functions under the condition of stationarity for the near-Sun solar wind turbulence for the first time based on measurements obtained from Parker Solar Probe (PSP) at 0.17 au. We find that the scaling indices $\xi$ of magnetic field show a linear dependence on the order $p$ close to $\xi(p)=p/4$. The multifractal scaling of magnetic-trace structure functions becomes monoscaling close to $\xi(p)=p/3$ with the local magnetic field perpendicular to the sampling direction and close to $\xi(p)=p/4$ with the local magnetic field parallel to the sampling direction when measured with the stationary background magnetic field. The scaling of velocity-trace structure functions has similar but less significant changes. The near-Sun solar wind turbulence displays different scaling anisotropies with the near-Earth solar wind turbulence, suggesting the evolution of the nonlinear interaction process during the solar wind expansion.   
\end{abstract} 

\section{Introduction} \label{sec:intro} 
Turbulence is ubiquitous in astrophysical systems, exhibiting a power law behaviour in the inertial range characterized by the absence of an intrinsic length scale. The inertial range scaling for hydrodynamic turbulence was first introduced as Kolmogorov (K41) phenomenology \citep{Kolmogorov1941ANSD}, which supposes the nonlinear interactions between velocity eddies. For MHD turbulence, Iroshnikov-Kraichnan (IK) phenomenology is proposed \citep{Iroshnikov1963AZh,Kraichnan1965PhFl}. IK phenomenology takes the coupling of velocity and magnetic field fluctuations into account and supposes that nonlinear interactions occur between Alfv\'{e}n wave packets. Both phenomenologies predicts that the scaling indices $\xi$ follow a linear relation with order $p$ (Kolmogorov scaling: $\xi=p/3$; IK scaling: $\xi=p/4$) under the idea of self-similarity and the assumption of homogeneity and isotropy.

Solar wind provides an ideal laboratory for the MHD turbulence. The inertial range scaling of solar wind turbulence was first investigated using Voyager 2 data \citep{Burlaga1991JGR,Burlaga1991GeoRL} and an multifractal scaling laws were found, indicating the existence of intermittency. Various models were proposed to account for the influence of intermittency on the scaling indices, including p-model \citep{Meneveau1987PhRvL}, $\beta$-model \citep{Marsch1993AnGeo}, bi-fractal model \citep{Ruzmaikin1995JGR}, and extended cascade model \citep{Tu1996AnGeo}. These models put efforts to recover the true scaling of the turbulent fluctuations underneath the intermittency. Both the Kolmogorov scaling or the IK scaling are obtained, leading to the difficulty to decide which phenomenology describes appropriately the solar wind turbulence due to the relatively small amount of data. Another approach to reveal the influence of intermittency is to compare the scaling indices before and after removing them. The magnetic field is found to exhibit a Kolmogorov scaling after removing intermittency using the widely used LIM (Local Intermittency Measure) technique \citep{Bruno2001PSS} to identify the intermittency \citep{Salem2009ApJ, Wang2014ApJ, Yang2017ApJ}. 


The anisotropies of the scaling laws in the solar wind turbulence at MHD scales are also widely investigated. Multiple models are constructed to describe the anisotropy of MHD turbulence, including 2D turbulence \citep{Shebalin1983JPP}, slab+2D model \citep{Zank1996JGR}, critical balance \citep{Goldreich1995ApJ}, dynamic alignment \citep{Boldyrev2006PhRvL}. The power spectral index anisotropy, which supports the critical balance theory \citep{Horbury2008PhRvL, Podesta2009ApJ}, disappears after removing intermittency \citep{Wang2014ApJ}. The higher-order scaling shows the anisotropy to be monoscaling observed parallelly and multifractal scaling observed perpendicularly \citep{Osman2014ApJL}, which also disappears after taking away the intermittency \citep{Pei2016JGR}. Magnetic fluctuations favor the isotropic Kolmogorov scaling, rather than IK scaling at 1 au, while the velocity is closer to the IK scaling. The difference between the magnetic field and velocity is one of the major puzzles in the study of solar wind turbulence.

However, the magnetic fluctuations in the solar wind consist of not only the turbulence, but also various structures possibly including current sheets \citep{Li2007ApJ, Miao2011AnGeo}, convective magnetic structures \citep{Tu1991AnGeo, Tu1993JGR}, tangential turnings corresponding to loops of magnetic field lines in the photosphere \citep{Nakagawa1989JGR, Tu2016AIP}, flux tubes originated at the sun \citep{Bruno2001PSS, Borovsky2008JGR}, interplanetary magnetic flux ropes \citep{Tu1997AnGeo,Zhao2020ApJ}, and magnetic holes \citep{Yu2021ApJ}. Fluctuations of these structures certainly should not be included in the turbulence scaling analyses. \cite{Wu2020ApJ} pointed it out that the convective structure and large variations may have an impact on the averaged magnetic fields, and the local mean magnetic field may not represent the major direction of the individual measurement. The stricter determination of sampling angle leads to different scaling indices, demonstrating the importance of a precise sampling direction for the study of anisotropy \citep{Wang2015ApJ, Telloni2019ApJ, Wu2022}. To avoid the effect of structures, \cite{Wu2020ApJ} proposed the criterion $\phi<10^\circ$ to select the time stationary subintervals such that the convective structures are rejected and their averaged magnetic field directions represent the major instantaneous field directions in the corresponding intervals. Using this criterion, \cite{Wu2020ApJ} found that the second-order scaling indices of both magnetic field and velocity are isotropic for near-Earth solar wind turbulence with a time stationary background field, and \cite{Yang2021ApJ} confirmed the effect of the large-scale field structures with a driven compressible three-dimensional MHD turbulence.  
  
Recently, Parker Solar Probe (PSP) provides an opportunity to analyze the near-Sun solar wind turbulence. The power spectra of magnetic field are reported to be close to the IK spectrum \citep{Chen2020ApJS} and the spectral index is close to $-5/3$ and $-3/2$ respectively when observed parallel and perpendicular to the local magnetic fields direction \citep{Huang2022ApJL}. The higher-order scaling presents nonlinear behaviours \citep{Alberti2020ApJ,Sioulas2022ApJ}. \cite{Wu2022arxiv} used the data from both PSP and Ulysses and analyzed the multi-order scalings anisotropies without considering the effect of non-stationarity. They found that the anisotropy of the multi-order scalings presented similar nonlinear behaviour for the two datasets and proposed the possibility that the inertial range contains two subranges. However, the effect of intermittent structures on the higher-order scalings and their anisotropy has not been considered in the near-Sun solar wind.

Here, we perform higher-order statistics and analyze the anisotropy of the higher-order scaling with considering the effect of stationarity for the first time on the near-Sun solar wind using PSP observations. We apply the criterion $\phi<10^\circ$ to ensure the time stationarity and demonstrate the influence of possible structures. We find that the scalings of the magnetic field and velocity are anisotropic. The scaling indices are close to the K41 prediction in the perpendicular sampling direction and close to IK prediction in the parallel sampling direction. This paper is organized as follows. In Section \ref{sec:DATA}, we describe the PSP measurements and the calculation of conditioned structure functions. In Section \ref{sec:RESULTS}, We show both the original and conditioned structure functions and their scaling indices. In Section \ref{sec:CONCLUSIONS}, we draw our conclusions.

\section{Data and Method} \label{sec:DATA}  
We use magnetic field data with time resolution of $0.8738$ second measured by the fluxgate magnetometer in the FIELDS instrument suite \citep{Bale2016SSRv} and plasma data from the the Solar Probe Cup \citep{Case2020ApJS} of Solar Wind Electrons, Protons, and Alphas instrument suite \citep{Kasper2016SSRv} on board PSP. One-day-long time interval is taken during the first encounter phase on 2018 November 6 when $r \sim 0.17 $ au, $V_\mathrm{0}= 332 $ km/s, and cross helicity $\sigma_c = 0.84$. During this interval, PSP observed the solar wind emerging from a small negative equatorial coronal hole \citep{Badman2020ApJS}.
 
The magnetic field increments are obtained by
\begin{equation}
\delta \vec{B}(t,\tau)=\vec{B}(t) -\vec{B}(t+\tau), 
\end{equation}    
where $\tau$ is the time lag, $\vec{B_\mathrm{1}}=\vec{B}(t)$ and $\vec{B_\mathrm{2}}=\vec{B}(t+\tau)$. We calculate the local background magnetic field $\vec{B_\mathrm{0}} (t,\tau)$ as averages in the moving scale-dependent window $\tau$. For each pair of magnetic field $\vec{B}(t)$ and $\vec{B}(t+\tau)$ separated by a time lag $\tau$, corresponding to increments $\delta \vec{B}(t,\tau)$, we also calculate three angles. The first $\theta_\mathrm{VB}(t,\tau)$ is the sampling angle between the local background magnetic field and the flow velocity measured in the spacecraft frame to take the spacecraft velocity of PSP into account \citep{Zhao2020ApJ, Duan2021ApJ}.  The second $\phi(t, \tau)$ is the angle between the two magnetic fields $\vec{B_1} = \left<\vec{B}(t)\right>|^{t}_{t+\tau/2}$ and $\vec{B_2} = \left<\vec{B}(t)\right>|^{t+\tau/2}_{t+\tau}$ , where $\left< \right>$ denotes an ensamble time average. $cos\ \phi(t, \tau)= \frac{\vec{B_1}\cdot\vec{B_2} }{|\vec{B_1}||\vec{B_1}|}$. $\phi$ weighs the time stationarity. The third $\Delta \theta (t,\tau)$ is the angle between $\vec{B}(t)$ and $\vec{B}(t+\tau)$, representing the angular change in the magnetic field direction.

The structure functions with $p$-th order $S^p (\tau)$ for the magnetic field are calculated by
\begin{equation}
S^q (\tau)= <|\delta \vec{B}(t,\tau)|^p>,
\end{equation}    
where $<>$ denotes an ensamble time average. $\delta \vec{B}(t,\tau)$ are selected under the criteria of different sampling angle ($80^\circ<\theta_\mathrm{VB}(t,\tau)<100^\circ$ and $\theta_\mathrm{VB}(t,\tau)<10^\circ\ or\ \theta_\mathrm{VB}(t,\tau)>170^\circ$) to obtain $S^p (\tau_{\perp})$ and $S^p (\tau_{\parallel})$. We perform power law fits within $10<\tau<100$ s and obtain the scaling indices $\xi(p)$, $\xi_{\perp}(p)$, $\xi_{\parallel} (p)$ from $S^p (\tau)$, $S^p (\tau_{\perp})$, and $S^p (\tau_{\parallel})$, respectively. These quantities are obtained for the velocity using the same procedure except that the magnetic field increments are replaced by the corresponding increments $\delta \vec{V}(t,\tau)=\vec{V}(t) -\vec{V}(t+\tau)$.

In order to truly determine the structure functions of turbulence and analyze their anisotropy, we use the criterion $\phi(t,\tau)<10^\circ$, which has been shown to be easy but effective to ensure time stationarity of the background magnetic field and avoid the effect of the magnetic structures \citep{Wu2020ApJ,Yang2021ApJ}. The ensamble time average are limited to those $\delta \vec{B}(t,\tau)$ and $\delta \vec{V}(t,\tau)$ with $\phi(t,\tau)<10^\circ$. We again perform power law fits within $10<\tau<100$ s and obtain $\xi(p)$, $\xi_{\perp}(p)$, $\xi_{\parallel} (p)$ for those conditioned structure functions $S_{\rm{R}}^p (\tau)$, $S_{\rm{R}}^p (\tau_{\perp})$, and $S_{\rm{R}}^p (\tau_{\parallel})$ with stationary background field.

\section{Results} \label{sec:RESULTS}  
\begin{figure}[ht!] 
\includegraphics[width=1.0\linewidth]{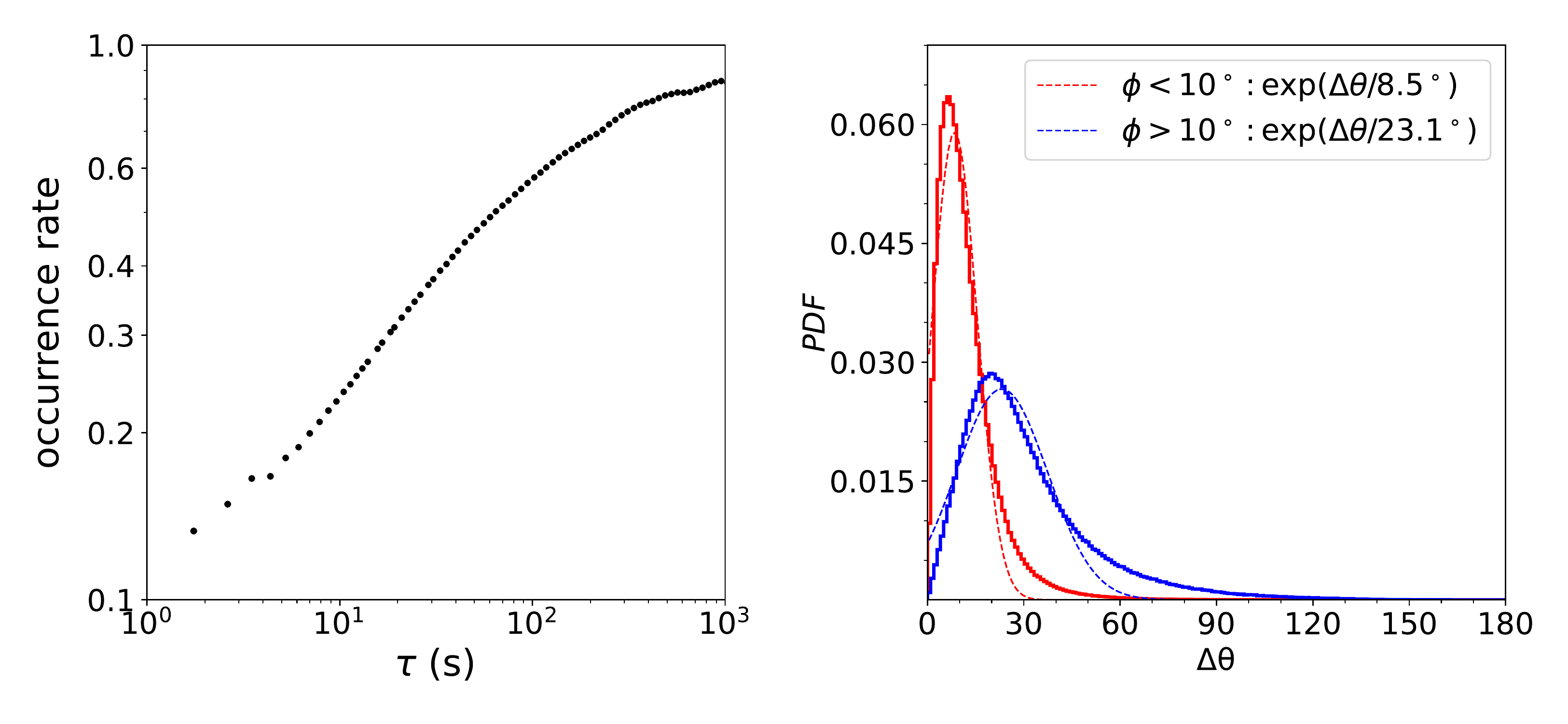} 
\caption{\label{fig:fig1} Left: occurrence rates of subintervals with $\phi>10^\circ$ as a function of time lags $\tau$. Right: the probability distribution functions of all angular change $\Delta \theta$ in the magnetic field direction within $10<\tau<100$ s. The red corresponds to the subintervals with $\phi<10^\circ$ and the blue with $\phi>10^\circ$. The dashed lines represent the exponential fits to the two portions, with the fitted parameters shown in the legends.}
\end{figure}

Fig. \ref{fig:fig1} (left) demonstrates the occurrence rate of the subintervals with $\phi>10^\circ$ as a function of time lags $\tau$. The occurrence rate increases from $\sim 0.2$ to $\sim 0.8$ as the time lag increases from 10 s to 1000 s. It reflects that the non-time-stationary measurements occupy a large part of the fluctuations and cannot be neglected.  In this study, we focus on the range from 10 s to 100 s, since the occurrence rates of non-stationarity are relatively high in this range. Fig. \ref{fig:fig1} (right) presents the probability distribution functions of the angular change $\Delta \theta$ with the time lag inside $10-100$ s under the conditions of $\phi<10^\circ$ and $\phi>10^\circ$, respectively. The exponential fits to the two portions (exp$(\Delta \theta/8.5^\circ)$ and exp$(\Delta \theta/23.1^\circ)$) show similarity to that obtained from ACE measurements for the near Earth solar wind: one is for the solar wind turbulence with exp$(\Delta \theta/9.4^\circ) $ and the other is the crossings of the interfaces between flux tubes with exp$(\Delta \theta/24.4^\circ) $ \citep{Bruno2001PSS,Borovsky2008JGR}, suggesting that those subintervals with $\phi>10^\circ$ may represent the magnetic walls between flux tubes. 
  
 \begin{figure}[ht!]
\includegraphics[width=1.0\linewidth]{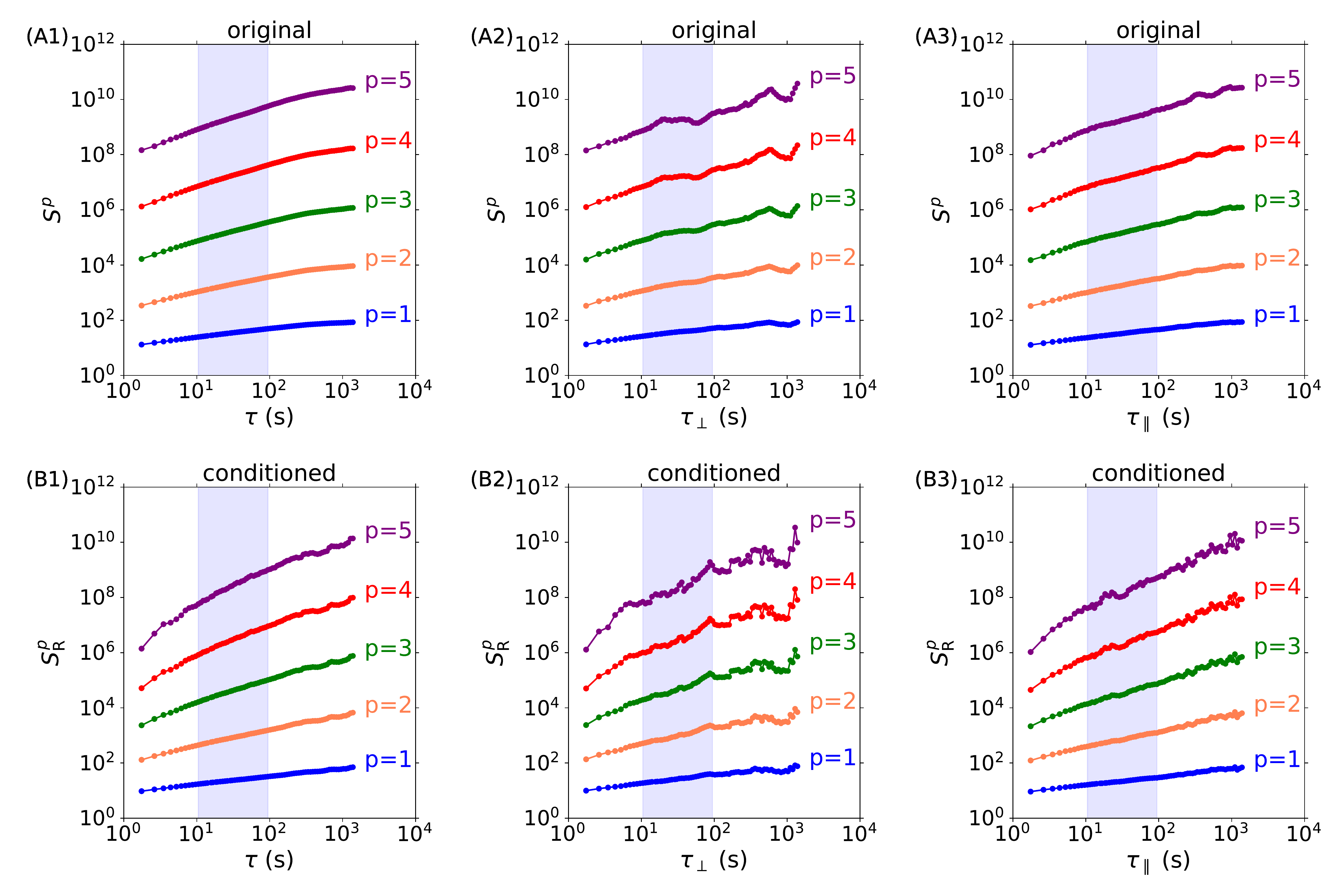} 
\caption{\label{fig:fig2} (A1-A3) The original magnetic-trace structure functions $S^p$ detected in the radial direction (A1), in the direction perpendicular (A2) and parallel (A3) to the local magnetic field. The blue, orange, green, red and purple represent the $S^p$ with order $p = 1, 2, 3, 4, 5$, respectively. The blue shadow denotes the range for the fitting indices. (B1-B3) The conditioned magnetic-trace structure functions $S_{\rm{R}}^p$ with stationary background with the same format as A1-A3.} 
\end{figure}

\begin{figure}[ht!] 
\includegraphics[width=1.0\linewidth]{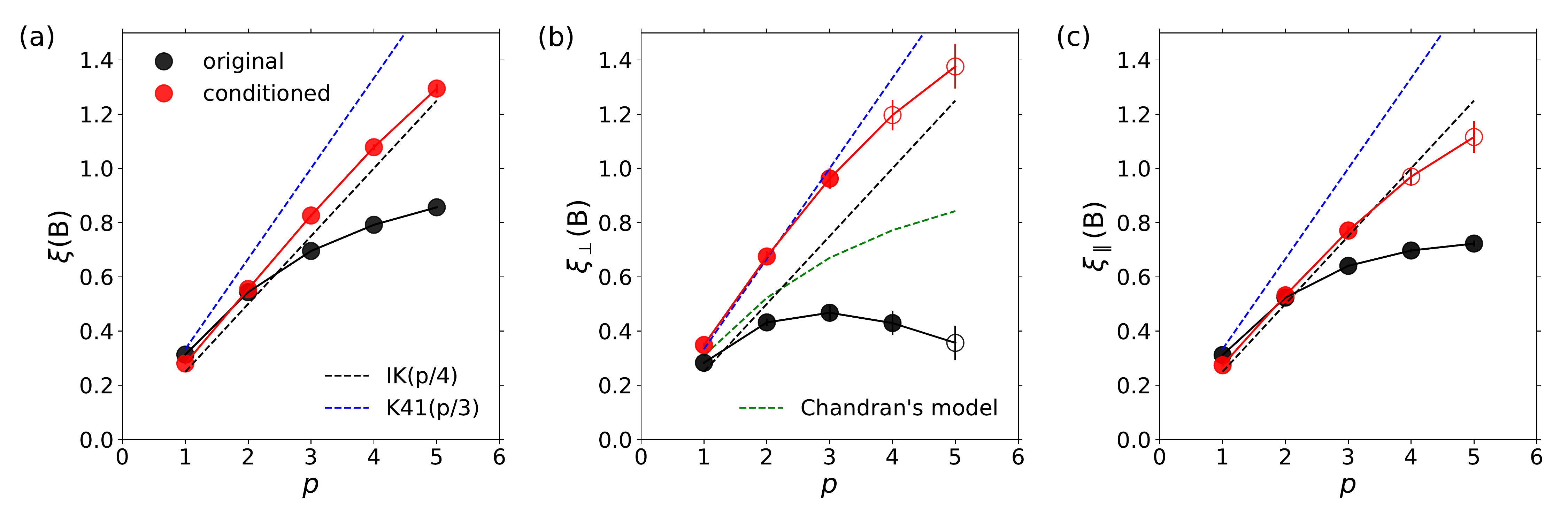} 
\caption{\label{fig:fig3} The scaling indices of magnetic-trace structure functions as a function of the order $p$ detected in the radial direction (a), in the direction perpendicular (b) and parallel (c) to the local magnetic field. The red and black represent the original indices and the indices with stationary background, respectively. The fitting errors are shown by the bars. Most of them are so small that they are covered by the points. The dashed black line denotes the IK predictions. The dashed blue line denotes the K41 predictions. The dashed green line denotes the predictions of \cite{Chandran2015ApJ}'s model. Those indices with higher than $p_{max}$ are marked in hollow style.
}
\end{figure}

 Fig. \ref{fig:fig2} presents $S^p (\tau)$, $S^p (\tau_{\perp})$, and $S^p (\tau_{\parallel})$ in the upper panels and $S_{\rm{R}}^p (\tau)$, $S_{\rm{R}}^p (\tau_{\perp})$, and $S_{\rm{R}}^p (\tau_{\parallel})$ in the bottom panels. The higher-order structure functions are less smooth for the perpendicular and parallel directions, which is reasonable since they are obtained from the subset of the samplings used to calculate the overall structure functions. The power-law behaviour of these structure functions over the range 10-100 seconds is obvious. The scaling indices are obtained through the least squares linear fitting in the logarithmic space from 10 seconds to 100 seconds and are shown in Fig. \ref{fig:fig3} with their corresponding standard errors. The scaling indices obtained from the original (conditioned) structures functions are displayed in black (red). In Fig. \ref{fig:fig3} (a), the nonlinear behaviour of the radial scaling indices are obvious. After avoiding the effect of the magnetic structures, the monoscaling becomes distinct as shown in red. The indices with first two orders barely change. The scaling indices are consistent with the IK scaling predictions.

We also present the scaling indices for the magnetic-trace structure functions observed in the direction perpendicular and parallel to the local magnetic field in Fig. \ref{fig:fig3} (b) and (c), respectively. The perpendicular scaling indices shown in Fig. \ref{fig:fig3} (b), change from multifratal scaling to approximate K41 monoscaling after the requirement of time stationary background field. Note that the highest trustworthy order $p_{max}$ is determined by the procedure in \cite{dewit2004PhRvE} and the indices with higher than $p_{max}$ are marked in hollow style. The multifratal scaling measured in the parallel direction also disappears, however, becomes close to the IK monoscaling. The anisotropy of the scaling is obvious under the condition of time stationary background field. 

 We also plot in green dashed line the prediction of the \cite{Chandran2015ApJ}'s model $\xi(p) = 1-\beta^p$ with $\beta=0.691$ calculated based on the assumption that the most intense coherent structures are two-dimensional in Fig. \ref{fig:fig3} (b). The observed scaling here for the perpendicular fluctuations deviates significantly from Chandran's prediction and the scaling indices present a decreasing trend with $p$. The nonlinear behaviour of the original scaling indices as a function of order $p$ shows that the higher order scaling indices are shallower than lower order scaling indices for the structure functions observed in the perpendicular to the local magnetic field. The decreasing trend of the scaling indices with $p$ was also found for the magnetic field \citep{Veltri1999AIP}, for the velocity \citep{Salem2009ApJ} and for the electric field \citep{Zhu2020ApJ}. These results are in contrast with the usual predicted monotonic increase \citep{Frisch1995}. \cite{He2019ApJ} constructed a unified paradigm to describe various intermittency-related quantities with the same set of parameters using Castaing function and found that the derivative ratio determines the order trend of the scaling indices. Here, we propose a possibility that the original decreasing scaling indices could result from the convective structures in the solar wind, which should be removed before the calculation. The linear trend of scaling indices should be recovered after the avoiding the effect of the intermittent structures.

\begin{figure}[ht!] 
\includegraphics[width=1.0\linewidth]{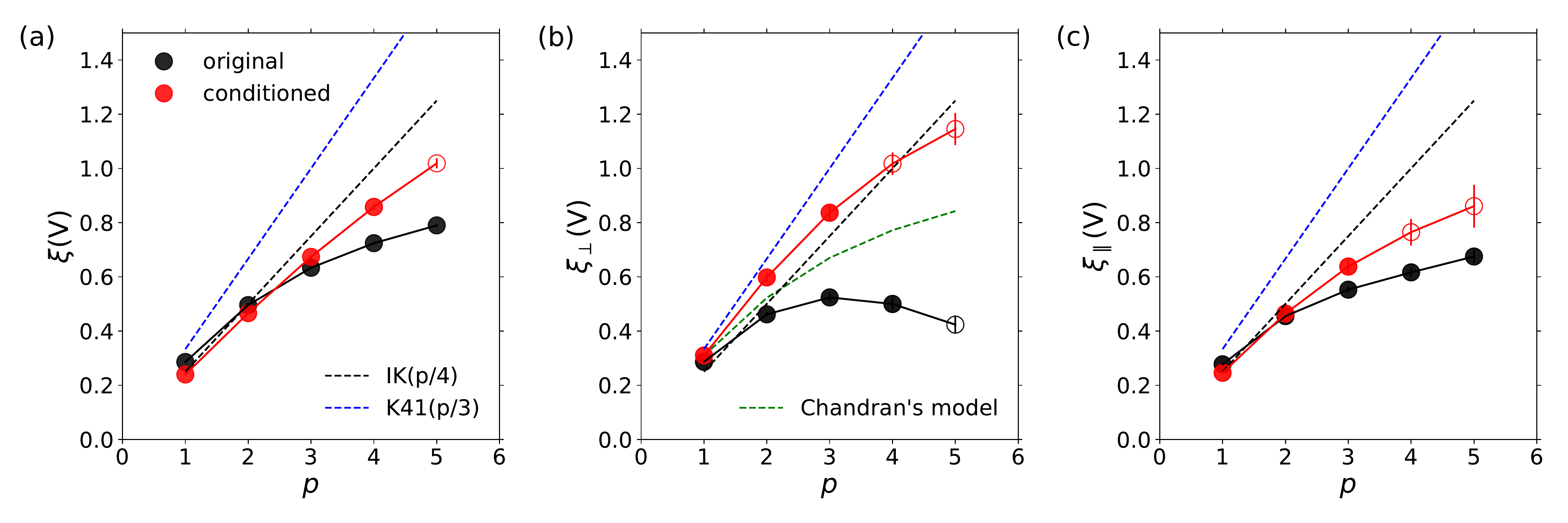} 
\caption{\label{fig:fig4} The scaling indices of velocity-trace structure functions as a function of the order $p$ with the same formats as in Fig. \ref{fig:fig3}.
}
\end{figure}

In Fig. \ref{fig:fig4}, we show the results for the velocity-trace structure functions. It is clear that the velocity exhibits a similar but less significant behaviour with the magnetic field. The multifratal scaling seems to disappear when the background field is required to be time stationary. The scaling indices for the velocity becomes close to the monoscaling measured both in parallel and perpendicular sampling direction.

The original perpendicular and parallel second-order indices are $0.43$ and $0.52$ for the magnetic field, which is different from previous studies \citep{Zhao2020ApJ, Huang2022ApJL} with different definition of sampling angles. The original perpendicular and parallel second-order indices are $0.46$ and $0.46$ for the velocity. After the stationarity criterion, the perpendicular and parallel second-order indices become $0.68$ and $0.53$ ($0.60$ and $0.46$) for the magnetic field (velocity). The anisotropies are revealed with a time stationary background field.

\begin{figure}[ht!] 
\includegraphics[width=1.0\linewidth]{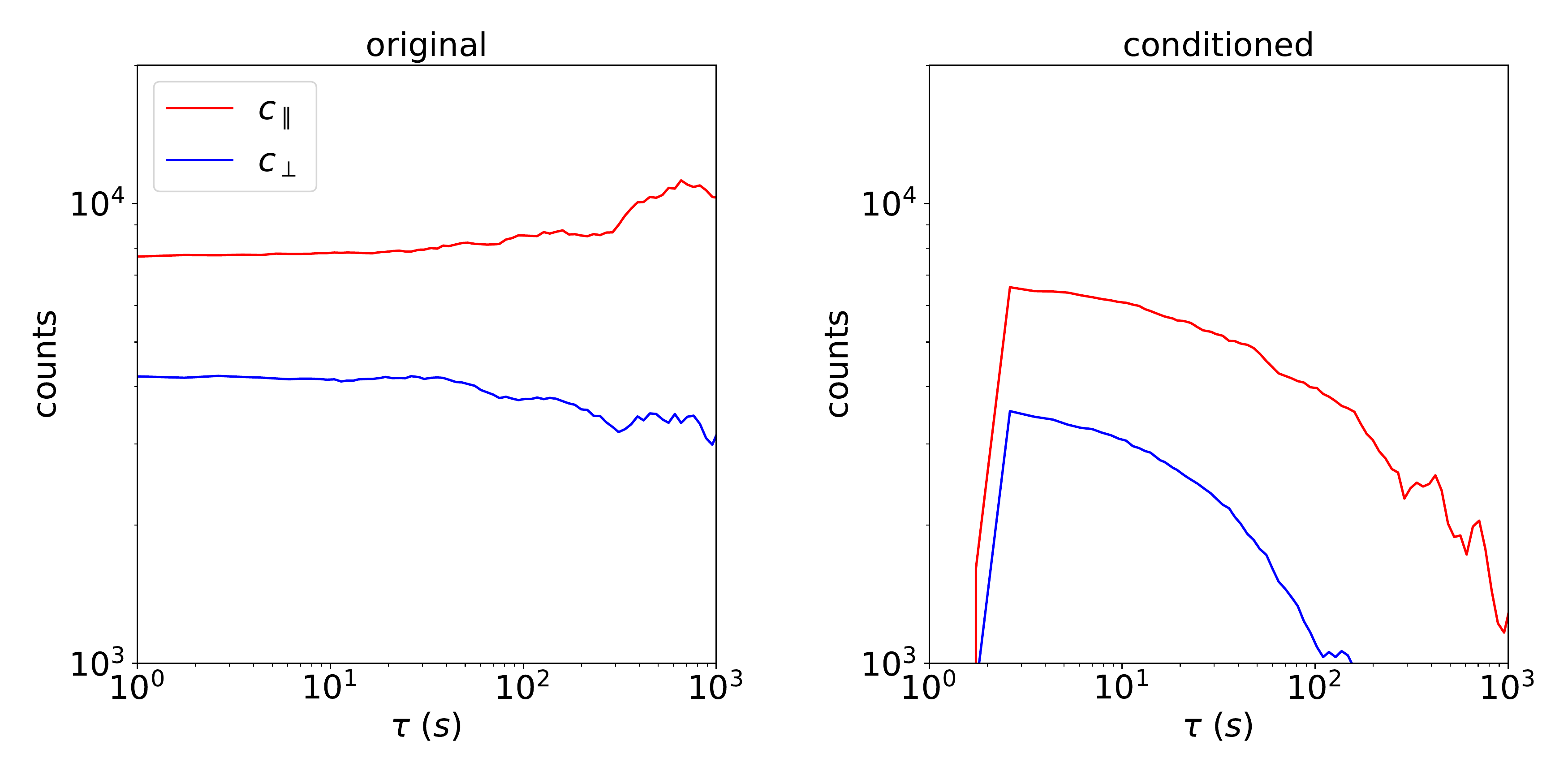} 
 \caption{\label{fig:fig5} The sampling counts as a function of time scale in the direction perpendicular (blue) and parallel (red) to the local magnetic field for the original (left) and conditioned (right) increments.
}
\end{figure}

 The overall structure functions are determined by both the power levels and the counts of the samplings in all directions. Fig. \ref{fig:fig5} displays both $c_\perp$ and $c_\parallel$, where $c_\perp$ and $c_\parallel$ are the counts of the sampled increments in the perpendicular and parallel directions respectively. It is clear that $c_\parallel$ is larger than $c_\perp$ for both original and conditioned increments, explaining why the scaling exponents in Fig. \ref{fig:fig3} (a) display a more similar behaviour with the parallel scaling exponents in Fig. \ref{fig:fig3} (c), instead of the perpendicular scaling exponents in Fig. \ref{fig:fig3} (b).

\section{conclusions and discussion} \label{sec:CONCLUSIONS}
In this Letter, we perform higher-order scaling analyses using the near-Sun observations, which is highly Alfv\'{e}nic slow solar wind coming from a small coronal hole. We apply the criterion of time stationarity of magnetic field in order to avoid the influence of magnetic structures. We find that the structures affect more strongly with increasing scale and their magnetic angular changes indicate they may be fossil flux tube walls originated from the Sun. We find that the scaling of the magnetic field and velocity are both approximately consistent with the IK monoscaling ($\xi=p/4$). We analyze the anisotropy of the higher-order scaling with considering the effect of stationarity for the first time and find that the multifractal scaling may result from the effect of magnetic structures. The scaling measured in the perpendicular direction is close to the K41 scaling ($\xi=p/3$) and the scaling measured in the parallel direction is close to the IK scaling ($\xi=p/4$). The scalings for the velocity show similar changes but the conditioned scalings are less closer to the monoscaling predictions. 

In a standard view, solar wind turbulence is described by two different sets of scaling laws, one (IK-like) in the direction of the magnetic field, and a different one (Kolmogorov-like) in the plane perpendicular to the magnetic field \citep{Marino2023PhR}. Our results are consistent with this standard view. The overall spectral index evolution from $-3/2$ in the near-Sun solar wind to $-5/3$ in the near-Earth \citep{Chen2020ApJS} may be a consequence of the combination of the anisotropy of the standard view and the measurement effect that PSP observed fluctuations mostly in the parallel direction in the near-Sun solar wind and more and more fluctuations in the perpendicular direction as solar wind expands.

The magnetic field data resampled to the plasma data cadence with a time resolution of 0.8738 seconds may suffer from the aliasing at high frequencies. In order to evaluate the effect of the aliasing, we also conduct a comparison with the case of un-resampled dataset with a time resolution of 1/292.969 seconds. The two power spectra of the magnetic field from both datasets (not shown) indicate the aliasing occurs at the frequency higher than 0.2 Hz and are overlapped in the analyzed region 0.01-0.1 Hz. We also calculated the scaling indices of the overall multi-order structure functions using the un-resampled dataset and confirm that they are the same as the resampled dataset in the region 0.01-0.1 Hz. Therefore, the resampling process of the magnetic field data does not affect our results.
  
Solar wind, as a natural laboratory, provide abundant clues for the understanding of intricate MHD turbulence, in which nonlinear interaction has always been one of the central issues. Solar wind turbulence, itself as a channel to transport energy, plays a critical role in heliophysics. Figuring out how it transfers energy is significant for modelling the heliophysics environment. Phenomenological approaches are effective ways but suffer from the various structures and solar wind conditions. We use a simple but effective time stationarity criterion to unveil the true scaling covered by structures and find that the near-Sun solar wind turbulence exhibits obvious anisotropy that the perpendicular scaling is close to the K41 scaling ($\xi=p/3$) and the parallel scaling is close to the IK scaling ($\xi=p/4$). This is different from the isotropic scaling features observed in the near-Earth solar wind turbulence \citep{Wu2020ApJ}. In future, it is of importance to obtain the true scalings in different solar wind conditions at different radial distances to interpret the evolution of solar wind turbulence.

\acknowledgments  
This work is supported by the National Natural Science Foundation of China under contract Nos. 42104152, 41974198,41925018,42174194,41874200,41874199 and  National Key R\&D Program of China No. 2021YFA0718600. X. Wang is also supported by the Fundamental Research Funds for the Central Universities of China (KG16152401, KG16159701).

\end{document}